\documentclass[a4paper,preprintnumbers,amsmath,amssymb,nofootinbib,10pt]{revtex4}

\usepackage[margin=2cm]{geometry}
\usepackage{amssymb}
\usepackage{mathrsfs}
\usepackage{graphics}
\usepackage{epsfig}
\usepackage{graphicx}
\usepackage{subfigure}
\usepackage{bm}

\def\beq{\begin{equation}}
\def\eeq{\end{equation}}
\def\baq{\begin{eqnarray}}
\def\eaq{\end{eqnarray}}

\def\fnl{f_{\rm NL}}

\def\gnl{g_{\rm NL}}

\def\taunl{\tau_{\rm NL}}
\def\hnl{h_{\rm NL}}

\def\k{{\bf k}}
\def\x{{\bf x}}
\def\y{{\bf y}}
\def\kh{k_{\rm hor}}
\def\kl{k_{\ell}}
\def\fnl{f_{\rm NL}}

\def\gnl{g_{\rm NL}}
\def\bea{\begin{eqnarray}}
\def\eea{\end{eqnarray}}
\def\be{\begin{equation}}
\def\ee{\end{equation}}

\begin{document}

\hfill{CERN-PH-TH/2011-282}\\
\phantom{}\hfill{NORDITA-2011-101}

\title{\LARGE Inhomogeneous non-Gaussianity}

\bigskip

\bigskip

\author{Christian T.~Byrnes,$^{\,a}$
  Sami Nurmi,$^{\,b\,}$
  Gianmassimo Tasinato,$^{\,c\,}$
 David Wands$^{\,c\,}$\\
 \hskip1cm\\
%\hskip1cm
%
%$^a$ Fakult{\"a}t f{\"u}r Physik, Universit{\"a}t Bielefeld, Postfach 100131, 33501 Bielefeld, Germany
$^a$CERN, PH-TH Division, CH-1211, Gen\`eve 23, Switzerland \\
$^b$NORDITA, SE-106 91, Stockholm, Sweden\\
$^c$Institute of Cosmology $\&$ Gravitation, University of Portsmouth, Dennis Sciama Building,\\\hskip0.2cm Portsmouth, PO1 3FX, United Kingdom
\\
\hskip1cm}
%E-mail: \email{cbyrnes@cern.ch}, \email{s.nurmi@thphys.uni-heidelberg.de}, \email{g.tasinato@thphys.uni-heidelberg.de}, \email{david.wands@port.ac.uk}}
%\date{\today}

\begin{abstract}

We propose a method to probe higher-order correlators of the
primordial density field through the inhomogeneity of local
non-Gaussian parameters, such as $\fnl$, measured within smaller
patches of the sky. Correlators between $n$-point functions measured
in one patch of the sky and $k$-point functions measured in another
patch depend upon the $(n+k)$-point functions over the entire sky.
The inhomogeneity of non-Gaussian parameters may be a feasible way to
detect or constrain higher- order correlators in local models of
non-Gaussianity, as well as to distinguish between single and
multiple-source scenarios for generating the primordial density
perturbation, and more generally to probe the details of inflationary
physics.

\end{abstract}

\maketitle

%\abstract{..}
%\begin{abstract}...
%\end{abstract}

\section{Introduction}

Primordial
non-Gaussianity is a window onto the early universe, which can be used to  distinguish between different models for
the origin of cosmological fluctuations \cite{Liguori:2010hx,Chen:2010xka,Byrnes:2010em,Komatsu:2010hc,Wands:2010af}. For this reason,
it is important to characterize as completely as possible the properties
of observable quantities in a non-Gaussian distribution.  In this paper, we discuss under what conditions local non-Gaussian observables can be inhomogeneous to a measurable level.

Consider subdividing the observable sky of size $H^{-1}$ into a
large number of small patches of size $\ell \ll H^{-1}$, and
measuring cosmological observables on each of these patches. The
effective values of the cosmological parameters can differ from one
patch to the other, since each small part of the sky experiences a
different local background. In particular primordial curvature
fluctuations within a small patch are defined with respect to a
local background which can be decomposed into a cosmological
background solution of the homogeneous field equations, plus the
cumulative effect of random modes with wavelengths larger than the
size $\ell$ of the patch (but smaller than the size $H^{-1}$ of the
observable sky).  Since different patches of our observable universe
are affected by different contributions from the random long
wavelength modes, each patch is defined on a different background
configuration. As a consequence, quantities characterizing  the
properties of $n$-point functions of curvature fluctuations can be
different in one patch or the other. Indeed the effect of
long-wavelength modes on the small-scale gravitational collapse of
dark matter halos is described by the peak-background split in
astrophysics and is used to explain the biased clustering of dark
matter halos \cite{Kaiser:1984sw}. In a non-Gaussian field the
long-wavelength modes also modulate the amplitude of the
short-wavelength density perturbations which can lead to a
distinctive scale-dependent bias which is  a powerful probe of
primordial non-Gaussianity \cite{Dalal:2007cu}. Of course there is
also a cosmic variance between patches, even if there are no
non-zero higher order correlators. However this effect will be small
as long as there are a large number of independent measurements  in every patch.

Correlators between $n$-point functions measured in one patch, and $k$-point functions measured in another patch, depend on $(n+k)$-point functions in the entire sky. As an example, we will show that the autocorrelation of $\fnl$ between
different patches depends on a  class of
parameters controlling 5- and 6-point functions, that  in the single-source case
reduce to
% consistency relations
% reduce these to
 squares of trispectrum parameters.
 %on the other.
  This implies that
inhomogeneities in quantities characterizing non-Gaussian
observables can be measurable, since at present we have only weak or
even no constraints on parameters controlling higher-order point
functions in the full sky.  Moreover, although the optimal way to
constrain higher $n$-point functions in the full sky would be
through a direct analysis of that correlator, in practice this is
exceedingly time consuming and computer intensive. This problem will
become significantly more pressing when the Planck satellite data
becomes avalailable in early 2013 \cite{planck}, since it has a much
higher resolution than the WMAP satellite \cite{Komatsu:2010fb}.
Only during the last few years have the first constraints on the
trispectrum been made
\cite{Desjacques:2009jb,Vielva:2009jz,Smidt:2010sv,Fergusson:2010gn,Smidt:2010ra},
and there are not yet any constraints on higher $n$-point functions.
Therefore,  our method of relating $n$-point functions to
inhomogeneities  of lower-order functions should also provide a
feasible and practical way of constraining higher-point functions in
the full sky, even though the method is  suboptimal. There is a
clear theoretical interest in constraining higher-order correlators:
they may be much larger than the correlators which have already been
constrained,  and their detection would provide important additional
information for characterizing the origin of primordial
fluctuations. Refs.~\cite{Sasaki:2006kq,Lin:2010ua,Suyama:2011qi}
consider the amplitude of correlators of higher order than the
trispectrum for local models of non-Gaussianity. See
\cite{Leblond:2010yq} for an analysis of  non-Gaussianity not
characterized by a local shape, that investigates 10-point functions
and beyond.

The inhomogeneities of non-Gaussian quantities that we analyze in
this paper are distinct from the anisotropies of inflationary
observables, arising when background vector fields (perhaps of
curvaton nature) are turned on
\cite{Dimastrogiovanni:2010sm,Bartolo:2011ee,Kanno:2008gn,Watanabe:2009ct,Watanabe:2010bu,Dulaney:2010sq,Gumrukcuoglu:2010yc,Dimopoulos:2011ws,Dimopoulos:2011pe}.
We are going to consider primordial non-Gaussianity due to scalar
fields that do not induce statistical anisotropy. Conversely,
generic models of inflation with vector fields may produce
statistical anisotropy but do not induce inhomogeneities of
inflationary parameters.
The inhomogeneities that we consider here are also different from the scale-dependence of the non-linearity parameters, that describe a variation with size rather than with position of patches \cite{Byrnes:2010ft}.

%{\bf CB pls check this new paragraph, its hard to concisely describe the difference between our work and Bartolo's + Kosta's. We also need to add some discussion of the relation to bias papers.} We stress that although we are considering the spatial variation of correlators in different patches, this is distinct from a scale-dependence of the non-linearity parameters which describes a variation with size rather than position of the patches \cite{Byrnes:2010ft}. We are also describing a different phenomena compared to models of an inhomogeneous $\fnl$ caused by a vector field (perhaps of a curvaton nature), see \cite{Dimastrogiovanni:2010sm,Bartolo:2011ee,Dimopoulos:2011ws,Dimopoulos:2011pe}. We are rather considering the inhomogeneity of correlators caused by larger but homogeneous higher order correlators.

The plan of this paper is as follows: We first focus on single-source scenarios, in which only one field generates the curvature perturbation. In Sec.~\ref{sec:1and2} we consider correlators of $\zeta$ and the power spectrum, in Sec.~\ref{sub-errors} of $\fnl$ and in Sec.~\ref{sec:1-vs-more} possible tests of single versus multi-source scenarios. In  Sec.~\ref{sec:multi} we focus on the more general formula of multi-source models, before going on to analyze in Sec.~\ref{sec:large}  when the inhomogeneities can be large enough to be observable. Finally we conclude in Sec.~\ref{sec:conclusions}.

  \smallskip

\section{Statistical Inhomogeneities in Single-Source Models}

Let us start by considering  single-source models where the curvature perturbation
$\zeta$ arises from fluctuations of a single scalar field. The single-source scenarios that we have in mind
do {\it not} necessarily correspond to single-field models of inflation. We have in mind  set-ups
 in which more than one scalar field may be present in the system, but only one of them, which
 we dub $\sigma$, is responsible for generating
the primordial curvature perturbation.
% All the other fields $\phi_i$ in the system are effectively homogeneous, and  affect
% only the background dynamics.
Examples are the curvaton model \cite{curvaton,LUW}, or the modulated reheating scenario \cite{Dvali:2003em,Kofman:2003nx}. These models are particularly interesting since they are capable of generating large non-Gaussianity of the local type,  potentially observable by the Planck satellite.

The statistics of the curvature perturbation in general depend on
the size of the patch we consider. This is due to long-wavelength
fluctuations which contribute to local background quantities in any patch
smaller than the horizon. The curvature perturbation within  a patch
of size $\ell$ is given by \cite{starob85,ss1,Sasaki:1998ug,lms,Lyth:2005fi}
  \beq
  \label{zetal}
  \zeta_{\ell}(\x)=
  N'(\sigma_\ell)\delta_\ell\sigma(\x)+\frac{1}{2}N''(\sigma_\ell)
  \delta_\ell\sigma(\x)^2+\frac{1}{6}N'''(\sigma_\ell)\delta_\ell\sigma(\x)^3+\ldots\ ,
  \eeq
where $\sigma_\ell$ denotes the background field value in the patch.
It consists of the classical homogenous solution $\bar{\sigma}(t)$,
and of fluctuations $\delta\sigma_{\k}$ with wavelengths greater
than the patch size, $k<a\, \ell^{-1}$ (here $a$ denotes the scale
factor). $N(\sigma)=\int H\,dt$ denotes the number of e-foldings
from an initial time $t_i$, soon after horizon exit during inflation of the modes of
interest, to some final time (for example, during the primordial
radiation-dominated era) at which $\zeta$ has frozen to its final,
constant value. The fluctuations $\delta_\ell\sigma(\x)$ include
modes ranging from $k=a\, \ell^{-1}$ to $k_{i}=a_i\,H_i$, the latter
 corresponding to modes that exit the horizon at the initial time
$t_i$~\footnote{Notice, however, that this cut-off does not lead to an additional dependence 
on $t_i$ of the quantity $\delta_\ell \sigma ({\bf x})$, besides the one already contained
in $\delta \sigma_{\bf k}$. Indeed, 
the upper 
limit $k_i$  is meant to characterize a UV cut-off. Pushing this cut-off to a slightly
 larger fixed scale, $k_{uv} > k_i$ would remove the additional dependence on $t_i$, without qualitatively changing 
 our results. This  since subhorizon modes are expected to provide only subdominant contributions to the evolution on superhorizon modes.} 
:
  \beq
  \delta_\ell\sigma(\x) = \int\limits_{a\,\ell^{-1} < k < k_{i}}\frac{{\rm
  d}\,{\k}'}{(2\pi)^3}e^{i\k'\cdot\x}\delta\sigma_{\k'}\ .
  \eeq
We use a top-hat window function to select the modes within this
momentum interval. Different choices of the window function should
not significantly alter our results. $\delta_\ell\sigma(\x)$
consists of superhorizon modes at the initial time, $\delta\sigma_{\k}$, which can subsequently be
treated as a classical, Gaussian random field with the two-point
function given by~\footnote{We assume that the scalar field $\sigma$ has
  canonical kinetic terms, and is characterized by slow-roll
dynamics during inflation.
  }
  \beq
  \langle\delta\sigma_{\k}\delta\sigma_{\k'}\rangle=(2\pi)^3\delta(\k+\k')\frac{H^2}{2k^3}\
  .
  \eeq
Here the brackets denote ensemble averages.

\begin{figure}[htbp]
  \begin{center}
    \resizebox{80mm}{!}{
    \includegraphics{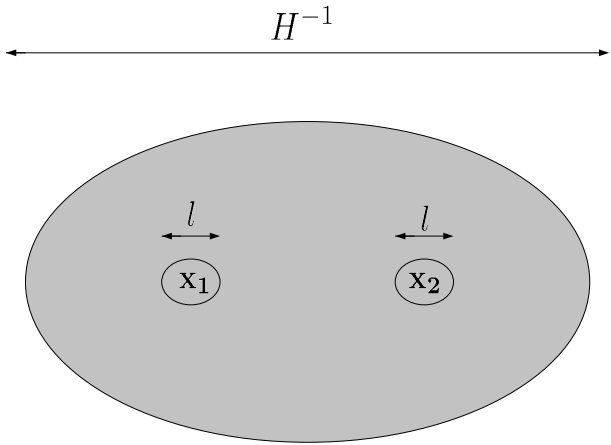}}
  \end{center}
  \caption{Schematic diagram to explain the measurements we are proposing. The large shaded area is the full sky (Hubble volume) while the two small regions of size $l\ll H^{-1}$ are two examples of the small patches in which observations are made and then correlated.}\label{figure}
\end{figure}

We are interested in computing $\zeta$ in patches smaller than the
observable universe, $\ell \ll H^{-1}$, see Fig~\ref{figure}.
%For definiteness, we assume
%the patches are spheres with radius $\ell / 2$.
The local background field
value in a patch centered at some position $\x$ is given by
  \beq
  \label{def_deltasigma}
  \sigma_{\ell, \,\x} = \sigma_{H^{-1}}+\int\limits_{a\,H< k < a\,\ell^{-1}}\frac{{\rm
  d}\,{\k}'}{(2\pi)^3}e^{i\k'\cdot\x}\delta\sigma_{\k'}\equiv
  \sigma_{H^{-1}}+\Delta_{\ell}\sigma_\x\ ,
  \eeq
where $\sigma_{H^{-1}}$ denotes the background field in the entire
observable universe. $\Delta_\ell\sigma_\x$ is comprised of fluctuations
$\delta\sigma_{\k}$ with wavelengths greater than the patch size,
$k<a\,\ell^{-1}$, which do not average out when computing spatial
averages over the patch $\ell.$ For a patch $\ell$ located at a
position $\x$, $\Delta_\ell\sigma_\x$ therefore acts as a constant, local
background.
 We emphasize that we use the label $\x$ to indicate quantities evaluated in a patch at position
 $\x$, and we consider them {\it not} as  functions of the coordinate $\x$.
 Denoting the spatial average over the patch by
$\langle\ldots\rangle_{\ell}$, we in general obtain the non-vanishing result
  \beq
  \Delta_\ell\sigma_\x\,=\,
  \langle\Delta_\ell\sigma_\x \rangle_{\ell}\, \neq \,0\
  ,
  \eeq
which depends on the location $\x$ of the patch.
On the other hand, when $\ell \ll H^{-1}$,  the spatial average of  $\Delta_\ell\sigma_\x$, when
  computed in the entire observable
universe $H^{-1}$, vanishes: $\langle\Delta_\ell\sigma_\x
\rangle_{H^{-1}}
%=\langle\Delta_\ell\sigma_\x \rangle
=0$.
This is because $\Delta_\ell\sigma_{\x}$ in our approximation depends
linearly on the fluctuations $\delta\sigma_{\k}$, which have a
vanishing spatial  average over the full sky. By the ergodic theorem,
and provided that the quantity $\ell\cdot H$ is sufficiently small,
the average of $\Delta_\ell\sigma_{\x}$ computed over the full sky
coincides with the ensemble average, $\langle\Delta_\ell\sigma_\x
\rangle_{H^{-1}}=\langle\Delta_\ell\sigma_\x \rangle=0$.

The consequences of long-wavelength fluctuations have been extensively studied in
the context of the infrared growth of inflationary correlators. It has been shown that, for
sufficiently local observations,
%,  for pure
% adiabatic fluctuations,
 a suitable  shift of background
quantities
 is able to remove infrared divergences
  associated with adiabatic field fluctuations
    \cite{Giddings:2010nc,Byrnes:2010yc,Gerstenlauer:2011ti,Urakawa:2010it},
    see also \cite{Linde:1994yf,Unruh:1998ic,Geshnizjani:2002wp}.
 On the other hand,  infrared-enhanced effects associated with  long wavelength
 isocurvature  fluctuations cannot be
 removed by shifts in background  quantities, and have the opportunity to
   provide sizable contributions
 to inflationary observables  \cite{Urakawa:2011fg}.
 %more than one field controls the background dynamics, as in the
%single source scenarios we have in mind, this is not possible and
%long-wavelength modes provide physical contributions
% to inflationary observables.
In our analysis, we will
 consider  correlators between cosmological observables
measured on distinct subhorizon patches within the observable
universe, in set-ups where
 %In such non-local observables the
  long-wavelength modes
generate sizable  statistical inhomogeneities. This effect was pointed out already
in \cite{Giddings:2011zd}, but  for the specific case of single-field slow-roll inflation where inhomogeneities are slow-roll suppressed.
Our treatment applies to generic single and multiple source models,
% that can be
% described by the $\delta N$ formalism
and  we find that the
inhomogeneities can become large in scenarios characterized by
observable non-Gaussianity. We will only consider the
inhomogeneities arising from scalar perturbations. Tensor modes,
included in the analysis of \cite{Giddings:2010nc,Gerstenlauer:2011ti},  should only
generate subleading, slow roll suppressed corrections to our results which we
neglect.

%-----------------------
%The long-wavelength fluctuations have been extensively studied in
%the context of infrared growth of inflationary correlators
%\cite{...}. It has been shown that, for pure single field models and
%for sufficiently local observations, a shift in the background
%quantities due to long-wavelength modes can be simply removed by a
%local coordinate rescaling  \cite{...}.
%%On the other hand,  when
%%more than one field controls the background dynamics, as in the
%%single source scenarios we have in mind, this is not possible and
%%long-wavelength modes provide physical contributions
%% to inflationary observables.
%In our analysis, we will
 %consider  correlators between cosmological observables
%measured on distinct subhorizon patches within the observable
%universe. In such non-local observables the long-wavelength modes
%generate statistical inhomogeneities. This was pointed out already
%in \cite{Giddings:2011zd} for the specific case of single-field slow
%roll inflation, where the effects however are slow roll suppressed.
%Our treatment applies for generic single and multiple field models,
%% that can be
%%described by the $\delta N$ formalism
%and  we find that the
%inhomogeneities can become very large in scenarios characterized by
%observable non-Gaussianity. We will only consider the
%inhomogeneities arising from scalar perturbations. Tensor modes,
%included in the analysis of \cite{Giddings:2011zd},  should only
%generate subleading, slow roll suppressed corrections to our results which we
%neglect.

After this general discussion aimed to define the set-up, we move on
discussing in more detail small scale statistical inhomogeneities of
quantities characterizing the properties of the curvature
perturbation.

\subsection{Correlators of one and two-point functions}\label{sec:1and2}

The two-point function of the curvature perturbation (\ref{zetal}) in a patch of size
$\ell\ll H^{-1}$ is given by
  \beq
  \langle\zeta_{\ell}(\y_1)\zeta_{\ell}(\y_2)\rangle_{\ell} =
  N'(\sigma_{\ell\,,\x})^2\langle\delta_{\ell}\sigma(\y_1)\delta_{\ell}\sigma(\y_2)\rangle_{\ell}+{\cal O}(\fnl^2\zeta^4)
  \eeq
If we re-express this with respect to the background field in the entire observable universe, $\sigma_{H^{-1}}$, using Eq.~(\ref{def_deltasigma}) this becomes
\beq
  \label{zetaL_2p}
  \langle\zeta_{\ell}(\y_1)\zeta_{\ell}(\y_2)\rangle_{\ell}
  = N'(\sigma_{H^{-1}})^2\langle\delta_{\ell}\sigma(\y_1)\delta_{\ell}\sigma(\y_2)\rangle_{\ell}
  \left(1+\frac{2N''(\sigma_{H^{-1}})}{N'(\sigma_{H^{-1}})}\,\Delta_{\ell}\sigma_\x\right)+{\cal O}(\fnl^2\zeta^4)
  \ ,
  \eeq
where $\x$ denotes the location of the patch and $\y_1,\y_2$ are
coordinates inside the patch. Recall that brackets $\langle\rangle_{\ell}$
denote spatial averages computed over the region $\ell$. The
long-wavelength contribution $\Delta_{\ell}\sigma_\x$ breaks
translational invariance: the two-point function depends not only on
the separation of the points, $|\y_1-\y_2|$, but also on the
location of the patch $\x$. The long-wavelength modes therefore
generate statistical inhomogeneity. For the case of 2-pt functions,
this fact was also pointed out in \cite{Giddings:2011zd} and, in
more generality,  in \cite{Lewis:2011au}.

Taking a Fourier transformation of (\ref{zetaL_2p}) within the
region $\ell$, i.e., integrating over the $\y$-coordinates and treating
$\x$ as a constant, we find a relation between the spectra of
$\zeta_{\ell}$ and $\zeta_{H^{-1}}$
  \beq
  \Delta_{\ell}{\cal P}_\zeta(\x)\equiv
  {\cal P}_{\zeta_\ell}-{\cal
  P}_{\zeta_{H^{-1}}}\,=\,\frac{12}{5}\,{\cal
  P}_{\zeta_{H^{-1}}}\,\fnl(\sigma_{H^{-1}})
  N'(\sigma_{H^{-1}})\Delta_{\ell}\sigma_\x\,
  +\,{\cal
  O}(\sqrt{\epsilon_{\sigma}}/N'{\cal P}_{\zeta}^{3/2})+{\cal
  O}(\fnl^2{\cal P}_{\zeta}^{2})\ ,
  \eeq
where the full sky power spectrum is
  \beq
  {\cal
  P}_{\zeta_{H^{-1}}}=N'(\sigma_{H^{-1}})^2\left(\frac{H}{2\pi}\right)^2\ ,
  \eeq
and $\fnl$ describes the amplitude of the primordial bispectrum relative to the square
 of the power spectrum, which is given in the $\delta N$-approach by \cite{Lyth:2005fi}
\beq
 \fnl = \frac56 \frac{N''}{N'^{2}} \,.
\eeq
$\Delta_{\ell}{\cal P}_\zeta(\x)$ measures deviations of the local spectrum,
measured in a patch of size $\ell$, from the global power spectrum characterizing
2-pt functions of
perturbations over the entire observable universe.

\begin{figure}[h!]
    \begin{center}
    \includegraphics[width=7 cm, height= 5.5 cm]{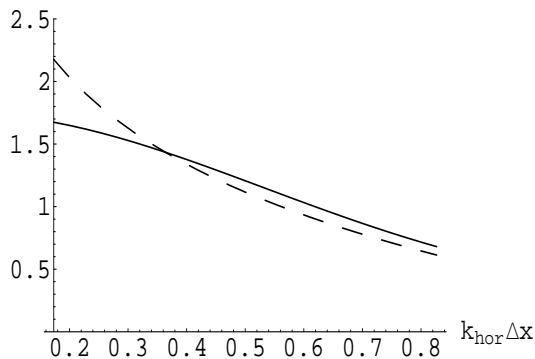}
    \caption{The exact integral Eq.
    \protect(\ref{F}) (solid line) and the approximation Eq.
    \protect(\ref{int_appr}) (dashed line) plotted for $\kh/\kl=\sqrt{0.03}$. For this choice, a patch of size $\ell$
    corresponds to a 3\% region of the observable sky, $H^{-1}$. On the $x$-axis we have plotted the values
    $\kl^{-1}<\Delta x<\kh^{-1}-\kl^{-1}$, corresponding to the possible distance between non-overlapping spheres
    of coordinate radius $\kl^{-1}/2$ within the observable universe.}
    \label{fig:comparison}
    \end{center}
  \end{figure}

Since $\Delta_{\ell}{\cal P}_{\zeta}(\x)$ is proportional to
$\Delta_{\ell}\sigma_\x$,  it is a Gaussian field with
 zero mean over the full sky for $\ell\ll H^{-1}$
$$\langle
\Delta_{\ell}{\cal P}_{\zeta}(\x) \rangle_{H^{-1}}\,=\,0\ ,$$
with two-point function
  \baq
  \label{dpzeta_2point_complete}
  \langle\Delta_{\ell}{\cal P}_{\zeta}(\x_1)\Delta_{\ell}{\cal P}_{\zeta}(\x_2)\rangle_{H^{-1}}
  &=&{\cal P}_{\zeta_{H^{-1}}}^2\left(\frac{12}{5}\fnl(\sigma_{H^{-1}})\right)^2N'(\sigma_{H^{-1}})^2
  \langle\Delta_{\ell}\sigma(\x_1)\Delta_{\ell}\sigma(\x_2)\rangle_{H^{-1}}\\\nonumber
  &=&{\cal P}_{\zeta_{H^{-1}}}^3\left(\frac{12}{5}\fnl(\sigma_{H^{-1}})\right)^2
  \int_{\kh}^{\kl}{\rm d}\,k'\frac{\,{\rm sin}(k' \Delta x)}
  {k'{}^2\Delta x}\ ,
  \eaq
where $\langle\ldots\rangle_{H^{-1}}$ denotes the spatial average computed
over the entire observable universe, which we assume coincides with
the ensemble average $\langle\ldots\rangle$. Hence we will drop the label
$H^{-1}$ from the angle brackets in what follows.
%and the angle brackets on the
%right hand side denote the ensemble average.
We  denote  $\Delta x = |\x_1-\x_2|$, $k_{\ell} = a\,\ell^{-1}$ and
$k_{\rm hor} = a\, H$. The integral can be evaluated as
  \baq
  F(\Delta x,\kh,\kl)&\equiv&
  \int_{\kh}^{\kl}{\rm d}\,k'\frac{\,{\rm sin}(k'\Delta x)}
  {k'{}^2 \Delta x}\nonumber \\
  &=&{\rm ci}(\kl\Delta x)-{\rm ci}(\kh\Delta x)-\frac{{\rm
  sin}(\kl\Delta x)}{\kl\Delta x}+\frac{{\rm
  sin}(\kh\Delta x)}{\kh\Delta x}\, , \label{F}
  \eaq
  with ${\rm ci}(x)$ the cosine integral function.
  Notice that, as expected, the correlators between two-point functions evaluated in different patches
 of the universe are proportional to $\left(\frac{6}{5}\fnl \right)^2\,=\,\tau_{\rm NL}$, which in
 single-source models is a parameter characterizing the four-point function measured in the entire sky.\footnote{However, as we will discuss in section \ref{sec:1-vs-more}, this tree level equality can be  broken when loop corrections are included.}
In the limit $\Delta x \to 0$,  the result takes a particularly simple form
  \beq
  F(\Delta x=0,\kh,\kl) = {\rm ln}\,\frac{\kl}{\kh}=-{\rm ln}\,(\ell H)\
  ,
  \eeq
corresponding to the well-known logarithm associated with
IR-enhancements.
 For $\kl^{-1} \ll \Delta x \ll \kh^{-1}$ it can be approximated
by
  \beq
  \label{int_appr}
  F(\Delta x,\kh,\kl)
  \approx 1-\gamma_{\rm E}-{\rm ln}(\kh\Delta x) + {\cal O}\left((\kl \Delta x)^{-1}\right)\
  \eeq
where $\gamma_{\rm E}\approx 0.58$ is the Euler  constant. A comparison
between the exact result (\ref{F}) and the estimate (\ref{int_appr})
is shown in Figure \ref{fig:comparison}.

The magnitude of the statistical inhomogeneities is described by the
variance of $\Delta_{\ell}{\cal P}_{\zeta}$. Using COBE normalization ${\cal
P}_{\zeta} = 2.4 \times 10^{-9}$ and the bound $|\fnl|\lesssim 10^{2}$ we find
  \beq
  \label{inhomogeneous_spectrum}
  \frac{\sqrt{\langle\Delta_{\ell}{\cal
  P}_{\zeta}(\x)^2\rangle}}{{\cal P}_{\zeta_{H^{-1}}}}=\frac{12}{5}{\cal
  P}_{\zeta_{H^{-1}}}^{1/2}\,|\fnl(\sigma_{H^{-1}})||{\rm ln}\,(\ell
  H)|^{1/2}
\,  \lesssim \, 10^{-2}\times|{\rm ln}\,(\ell H)|^{1/2}\ .
  \eeq
For patches $\ell$ corresponding to a few percent fraction of the
observable universe, the logarithm contributes  with a factor of
order unity, $|{\rm ln}\,(\ell H)| \simeq {\cal O}(1)$. \footnote{
Here and in the rest of  this paper, we will always  focus on cases
in which the size of the logarithms is of order one. The effects we
are interested in are observable  thanks to the coefficients in
front of the logarithms, that can assume large values.}
  The statistical
inhomogeneity seen on these scales could therefore be at a few
percent level. This observation was already made in
\cite{Giddings:2011zd} (see also \cite{arXiv:0806.0377}, and the more general discussion of
 \cite{Lewis:2011au}). Observational constraints were discussed in \cite{arXiv:0908.0963}.
 %Planck is expected to probe deviations from
%statistical isotropy down to $2$\% level and $21$ cm experiments
%could ultimately reach $10^{-7}$ sensitivity \cite{Pullen:2007tu}.

\smallskip

It is also interesting to consider the correlators of the quantity
$\Delta_\ell \zeta(\x)\equiv \zeta_\ell (\x)-\zeta_{H^{-1}} (\x)$,
measuring the difference in the amplitudes of the curvature
perturbation evaluated in the small patch $\ell$ and in the full sky
$H^{-1}$. To first order in perturbations we find
  \baq
  \Delta_\ell \zeta(\x) &=& \left(N(\x)-\langle N(\x)\rangle_{\ell}\right)-\left(N(\x)-\langle N(\x)\rangle_{H^{-1}}\right)\\
  \nonumber
  &=&\langle N(\x)\rangle_{H^{-1}}-\langle N(\x)\rangle_{\ell}\\\nonumber
  &=&-N'(\sigma_{H^{-1}})\Delta_{\ell}\sigma_{\x}\ .
  \eaq
The correlator between $\Delta_\ell \zeta(\x)$'s evaluated on different patches reads
  \be \langle \Delta_\ell
  \zeta(\x_1) \,\Delta_\ell \zeta(\x_2) \rangle_{H^{-1}}\,=\,{\cal
  P}_{\zeta_{H^{-1}}}\,F(\Delta x, k_{\rm hor}, k_\ell)
  \ee
  that is
proportional to the two point function measured in the full sky.
Also, we have \be
 \langle\Delta_\ell \zeta(\x_1) \Delta_\ell\ln {\cal P}_\zeta(\x_2) \rangle_{H^{-1}}\,=\, - \frac{12}{5}\,f_{\rm NL}\, {\cal P}_{\zeta_{H^{-1}}}\,F(\Delta x, k_{\rm hor}, k_\ell)
\ee
that is proportional to the three point function.

%Interestingly,  as done for the power spectrum it is also possible
%to define the quantity $\Delta_\ell \zeta(\x)$, and consider
%correlators among the long wavelength contributions to curvature
%perturbation evaluated in each different patch. The first term in
%the expansion of curvature perturbation within the small patch,
%using $\delta N$ formalism, is
%\bea \zeta_\ell (\x)&=& N\left(
%\sigma_{H^{-1}} + \Delta_\ell \sigma_\x + \delta \sigma_{\ell}(\x)
%\right)-\langle N \rangle_{\ell}
%\\
%&=&N' \left( \sigma_{H^{-1}} \right)\,\Delta_\ell \sigma_\x% - \langle
%%\Delta_\ell \sigma_\x\rangle_\ell\right)\,
%\,+\, N' \left( \sigma_{H^{-1}} \right)\,\delta \sigma_{\ell}(\x) \,+\,
%\cdots\ . \eea
%so we can write $\Delta_\ell \zeta(\x)\,\equiv\,
%\zeta_\ell (\x)-\zeta_{H^{-1}} (\x) $, and evaluate as above the
%correlator of this quantity in the full sky: \be \langle \Delta_\ell
%\zeta(\x_1) \,\Delta_\ell \zeta(\x_2) \rangle\,=\,{\cal
%P}_{\zeta_{H^{-1}}}\,F(\Delta x, k_{hor}, k_\ell) \ee that is
%proportional to the two point function as measured in the full sky.
%Also, we have \be
% \langle\Delta_\ell \zeta(\x_1) \Delta_\ell\ln P_\zeta(\x_2) \rangle\,=\, \frac{12}{5}\,f_{\rm NL}\, {\cal P}_{\zeta_{H^{-1}}}\,F(\Delta x, k_{hor}, k_\ell)
%\ee
%that is proportional to the three point function.

\smallskip

%Notice that in all our discussion we focus on considering correlators between quantities evaluated in different patches,
%and we do not discuss three and higher point functions. This since these are  proportional to higher
%powers of the power spectrum and are then suppressed
In this paper we will focus on the two-point correlators of
$n$-point functions measured in separate patches. It would also be
possible to extend our analysis to higher point correlators of the
same quantities  between different patches. This would require
extending our analysis to non-linear orders in $\Delta_\ell\sigma_{{\bf
x}}$ and, depending on the model, also extending the definition of
$\Delta_\ell\sigma_{{\bf x}}$, given in Eq. (\ref{def_deltasigma}), to
non-linear orders in perturbations.
%This would require to
%extend our definition of $\Delta_l\sigma_{{\bf x}}$, given in Eq.
%(\ref{def_deltasigma}), to non-linear order.
% thereby defining more observables.
 This should be technically straightforward although it would lead to longer expressions. We will not pursue this here, also since
 the results will be suppressed by higher powers of the power spectrum. Measuring the higher-order correlators of observables in different patches would also become computationally more demanding, something which our analysis, constraining the higher-point correlators, alleviates.

\subsection{Correlators of three-point functions}\label{sub-errors}

Because inhomogeneity of the power spectrum is already restricted by observational bounds on the primordial bispectrum,
it is more interesting to study the correlators of non-Gaussian observables such as $\fnl$ itself.
The non-linearity parameter $\fnl$ in a patch $\ell$, located at $\x$,
can be expressed as
  \beq
  \label{fnl_expand}
  \fnl(\sigma_{\ell},\x) =
  \fnl(\sigma_{H^{-1}})+\fnl'(\sigma_{H^{-1}})\Delta_{\ell}\sigma_\x+{\cal O}(\fnl''\Delta_{\ell}\sigma_{\x}^2)\ .
  \eeq
The prime denotes a derivative with respect to $\sigma_{H^{-1}}$ and for single-source models we find
  \beq
  \fnl'(\sigma_{H^{-1}}) = \left(\frac{9}{5}\gnl(\sigma_{H^{-1}})-\frac{12}{5}\fnl^2(\sigma_{H^{-1}})\right)N'(\sigma_{H^{-1}})\,,
  \eeq
where $\gnl$ is a parameter describing the primordial trispectrum relative to the power spectrum cubed, and is given in the $\delta N$-approach by \cite{Byrnes:2006vq}
\beq
\gnl = \frac{25}{54}\frac{N'''}{N^{\prime3}} \,.
\eeq
For brevity we suppress the argument $\sigma_{H^{-1}}$ in what
follows,  and write $\fnl(\sigma_{H^{-1}})\equiv \fnl$ etc meaning that quantities
without argument are evaluated in the full sky.

To leading order in $\Delta_{\ell}\sigma_{\x}$, the difference between $\fnl$
measured in the entire observable universe $H^{-1}$ and in a patch
of size $\ell$ is then given by
  \beq
  \Delta_{\ell}\fnl(\x)\equiv\fnl(\x)-
  \fnl= \left(\frac{9}{5}\gnl-\frac{12}{5}\fnl^2\right)N'
  \Delta_{\ell}\sigma_\x
  \ ,
  \eeq
and the two-point function of $\Delta_{\ell}\fnl(\x)$ reads
  \baq
  \label{dfnl_2point_complete}
  \langle\Delta_{\ell}\fnl(\x_1)\Delta_{\ell}\fnl(\x_2)\rangle
  &=&\left(\frac{9}{5}\gnl-\frac{12}{5}\fnl^2\right)^2
  N'{}^2\langle\Delta_{\ell}\sigma_{\x_1} \Delta_{\ell}\sigma_{\x_2}\rangle\\\nonumber
  &=&\left(\frac{9}{5}\gnl-\frac{12}{5}\fnl^2\right)^2
  N'{}^2\left(\frac{H}{2\pi}\right)^2\int_{\kh}^{\kl}{\rm d}\,k'\frac{\,{\rm sin}(k' \Delta x)}
  {k'{}^2\Delta x}
  \\\nonumber
  &=&\left(\frac{9}{5}\gnl-\frac{12}{5}\fnl^2\right)^2
  {\cal P}_{\zeta}\,F(\Delta x,\kh,\kl)\ .
  \eaq
We then learn that
$ \langle\Delta_{\ell}\fnl^2\rangle$ is proportional to a combination of terms containing
powers of $\gnl$ and $\fnl$. As we will discuss in the next sections, those terms
are related, by consistency relations valid in the single-source limit, to parameters controlling five
and six point functions.

 Using the amplitude of spectrum of curvature perturbations, ${\cal
P}_{\zeta} = 2.4 \times 10^{-9}$, we find
  \beq
  \label{dfnl_2point_corrections}
  \langle\Delta_{\ell}\fnl(\x_1)\Delta_{\ell}\fnl(\x_2)\rangle=7.8\times 10^{-9}\gnl^2 F(\Delta x,\kh,\kl)+{\cal O}(10^{-8}
  \fnl^2\gnl)+{\cal O}(10^{-8}
  \fnl^4)\ .
  \eeq
For $|\gnl| \lesssim \fnl^2$ the inhomogeneities are unobservably
small, $\langle(\Delta_{\ell}\fnl)^2\rangle\lesssim 1$, as $\fnl$ is
constrained by $|\fnl|\lesssim 10^2$ \cite{Komatsu:2010fb}. On the
other hand, if $|\gnl| \gg \fnl^2$, we discuss this possibility further in the next subsection, the
inhomogeneities can become significant as the observational
constraint for $\gnl$ is rather weak, $|\gnl|\lesssim 10^{6}$
\cite{Desjacques:2009jb,Vielva:2009jz,Smidt:2010sv,Fergusson:2010gn,Smidt:2010ra}.
In this limit, the last two terms in eq.
(\ref{dfnl_2point_corrections}), proportional to powers of $\fnl$,
give only subleading contributions and they can therefore be
neglected. The function $F(\Delta x,\kh,\kl)$ contributes with a
factor of order unity, and thus we arrive at the  result
  \beq
  \label{dfnl_2point}
  \langle\Delta_{\ell}\fnl(\x_1)\Delta_{\ell}\fnl(\x_2)\rangle \approx 10^{-8} \gnl^2 \lesssim
  10^{4}\ .
  \eeq
For $|\gnl|\sim 10^{6}$,  the variation of $\fnl$ measured in
different patches of size $\ell<H^{-1}$ becomes quite large,
$\langle(\Delta_{\ell}\fnl)^2\rangle\sim 10^{4}$, implying that
$\fnl$ becomes inhomogeneous on small scales~\footnote{\label{N''''}
We emphasize that our results are derived retaining only the linear
term in (\ref{fnl_expand}). For $|\gnl|\gg \fnl^2$, which is the
case of interest here, the higher order terms ${\cal
O}(\fnl''\Delta_{\ell}\sigma_{\x}^2)$ give a subleading contribution
provided that $|N^{''''}/N'{}^4|\ll 10^{5}|\gnl|$.}. This is a
generic feature of all single-source models with large $\gnl$.
\footnote{A comparable constraint on the inhomogeneity of $\fnl$ can
be found by considering the correlator of $\Delta_{\ell}\fnl$ and
$\Delta_{\ell}\zeta$, see (\ref{zetafnl}) for this correlator in the
multiple-source case. We thank Antony Lewis for pointing this out.}

It is illuminating to compare the result with the observational
accuracy for detecting $\fnl$ using a small fraction of the sky. For
example, the BOOMERanG experiment found $-670 < \fnl < 30$ at $65$\%
CL observing a $3$\% region of the sky \cite{Natoli:2009wk}. For
patches of the same size, we find the variance
$\langle(\Delta_{\ell}\fnl)^2\rangle \approx 1.4\times
10^{-8}\gnl^2$ using Eq. (\ref{dfnl_2point_corrections}). For
$|\gnl| \sim 10^{6}$, the variation of $\fnl$ measured on different
3\%-of-the-sky patches is therefore of the same order of magnitude
as the BOOMERanG $1$-$\sigma$ accuracy for detecting $\fnl$.

In general, when estimating $n$-point functions on a limited part of the
  sky, observational errors increase  (at least) proportionally to $1/\sqrt{f_{\rm sky}}$ ($f_{\rm sky}$ denotes the fraction
  of the sky which is observed), since there is less data available
  in a small patch.  Therefore when dividing the sky into, say,  one
  hundred  equal patches, the error bar on every observation will grow by at least a factor of ten compared to the full sky measurement. On the other hand, we then correlate, over the full sky,
  the measurements of $n$ and $k$-point functions   evaluated
  in distinct  small patches. This implies that, for the aim of constraining the values
  of  higher-order point functions in the full sky,
   we are able  to  regain much of the accuracy we had  previously lost.
   %  on characterizing the full sky $(n+k)$-point functions, since the data
  % set increases. Consequently, our method allows to constraint higher order point functions in the full sky
  % with  good accuracy.

There is however an important second effect to take into account:
when making measurements of  $n$-point functions in a patch smaller
than the entire sky,  a smaller number of correlators can be built.
For example,  for $n=3$, when estimating $\fnl$ in a patch the
range of scales available is  reduced, thereby limiting our ability
to analyze the squeezed limit of triangles~\footnote{We thank Rob
Crittenden and Dominic Galliano for  discussions on this point.}.
This effect is expected to depend only logarithmically on the size
of the patches. Indeed, the signal to noise ratio for measurements
of local non-Gaussianity scales as
\cite{Babich:2004yc,Riotto:2010nh}
  \beq
  \left(\frac{S}{N}\right)^2\propto \ell_{\rm max}^2\left(1-\frac{\ell_{\rm min}^2}{\ell_{\rm max}^2}\right)
  {\rm ln}\left(\frac{\ell_{\rm max}}{\ell_{\rm min}}\right)\ ,
  \eeq
where $\ell_{\rm max}$ and $\ell_{\rm min}$ denote the largest and
smallest multipoles accessible by the experiment. For example, for patches
covering one per cents of the sky, $\ell_{\rm min}/\ell_{\rm
max}\ll 1$, and the patch size, which determines $\ell_{\rm min}$,
enters only in the logarithm, reducing the accuracy  only by a factor of order three
with respect to full sky. We will not develop this issue further
in our work, but to account for these  logarithmic effects we simply
increase by one order of magnitude the values of our expected
constraints on higher order $n$-point functions in the full sky. We
will see that, even within this conservative estimate, we can find
interesting constraints on these quantities.

\smallskip

To conclude this section,
let us point out that alternative techniques based on needlets analysis of CMB data are particularly
well suited for testing  non-Gaussianity in selected regions of the sky.
% as needed for
%testing inhomogeneities.
This fact has already been used for investigating  inhomogeneity of non-Gaussianity
in different,   large  regions of the sky in \cite{Rudjord:2009au,Pietrobon:2009qg},
 with the main aim of investigating foreground
contaminations and directional variations of $\fnl$. Although no significant hints of anisotropies
have been reported in those studies,
it would be very interesting to apply the same techniques
to a collection of smaller regions of the sky, to instead test inhomogeneities of $\fnl$ along the lines we suggest here.

\subsection{New perspectives for discriminating single from multiple source models}\label{sec:1-vs-more}

We have learned from the previous analysis that a large $\gnl$ leads
to sizable inhomogeneities of $\fnl$. On the other hand,  other
inflationary observables can be affected by a  particularly large
$\gnl$. We discuss in this section this possibility, showing that the
conditions allowing to obtain large inhomogeneities for $\fnl$ offer
  new perspectives
 for distinguishing  single from multiple-source scenarios.

Loop corrections \cite{Seery:2010kh} are known to influence
observables by providing logarithmic contributions that in some scenarios can be large enough to dominate over tree-level quantities \cite{Suyama:2008nt,Kumar:2009ge}. In our set-up, $\taunl$ is  the observable that is most
sensitive to loop  corrections when $\gnl$ is large. Expressing it as
$$
\taunl\,=\, \taunl^{\rm tree}+ \taunl^{1-\rm loop}\,,
$$
one finds that
%It is also interesting to note that
%a large $\gnl$ generates a
%sizeable $\taunl$ through loop corrections. For
for  $|\gnl|\gg
|\fnl|^2$, the dominant part of the one-loop correction to $\taunl$
is given by
   \beq
   \label{taunl_loop_single}
   \taunl^{1-\rm loop}\, \approx\, \left(\frac{54}{25}\right)^2
   \,\gnl^2\,{\cal P}_{\zeta}\,\,{\rm ln (k\ell)}\,\lesssim\,
   10^4\ ,
   \eeq
assuming that the fourth order derivative $N''''$ is constrained
according to footnote \ref{N''''}. We have used approximation methods
similar to \cite{Boubekeur:2005fj,Lyth:2007jh,Bramante:2011zr} in
evaluating the loop integral. The cut-off scale $l$ should be a bit larger than the box in which we are making our measurements, such that $\ln(k\ell)\gtrsim 1$: because we are making a leading log approximation one cannot consider the limit of the logarithm going to zero. Comparing equations
(\ref{dfnl_2point_complete}) and (\ref{taunl_loop_single}), we
observe that the inhomogeneities of $\fnl$ are directly related to
the fraction of $\taunl$ generated by loops
%{\bf GT I added a factor 1/2 in the RHS, putting together  the numerical factors in the previous formulae.}
%  \beq\label{inlop}
%  \langle\Delta_{\ell}\fnl(\x_1)\Delta_{\ell}\fnl(\x_2)\rangle
%   \simeq \left(\frac{5}{6}\right)^2 \,\taunl^{1-{\rm loop}}\ .
%  \eeq
   \beq\label{inlop}
   \langle(\Delta_{\ell}\fnl)^2\rangle \simeq \left(\frac{5}{6}\right)^2 \,\taunl^{1-{\rm loop}}\ .
   \eeq
This consistency relation, which holds for $|\gnl|\gg |\fnl|^2$, is
specific for the single-source case. As we will learn in Sec
\ref{sub_afnl},
 in multiple source models
there are additional quantities contributing to the right hand side of Eq. (\ref{inlop}),
 which in principle make it possible to discriminate between
single and multiple source scenarios.
%A similar probe is given by
%the relation between $\gnl$ and $\taunl$ generated by loops
%({\ref{taunl_loop_single}), which is specific for single-source
%models.

Although for some models one cannot realise $|\gnl|\gg\fnl^2$, especially for those with quadratic potentials, for a review see \cite{Suyama:2010uj}, it is certainly possible to realise $|\gnl|\gg\fnl^2$ in general early universe models. This can happen for example in
the interacting curvaton scenario \cite{Enqvist:2009ww}, since the value of $\fnl$ oscillates depending on the initial field value of the curvaton, and the points where $\fnl=0$ do not coincide with points where $\gnl$ is suppressed.
Also an isocurvature field direction during inflation with a quartic
potential and zero VEV gives rise to $\gnl$ but not $\fnl$
\cite{Bernardeau:2002jf}.
 In general finite volume effects will also generate a non-zero $\fnl$
in
horizon size patches \cite{Bernardeau:2003nw,Bernardeau:2007xi},
 but in some patches $\fnl$ will still be small while $\gnl$ is not.
 Some models of multifield inflation can also give rise to $|\gnl|\gg\fnl^2$ at the end of inflation, see Eq.~(82) of \cite{Byrnes:2009qy} (see also \cite{Battefeld:2009ym}). 

Interestingly, large loop corrections to $\taunl$  also break the
well-known consistency relation $\taunl = (6\fnl/5)^2$,
characteristic for single-source models. This relation is indeed
{\it not} protected against loop corrections, as can be nicely
demonstrated by considering models where the curvature perturbation
takes the form $\zeta = N'\delta\phi + N'''\delta\phi^3/6$, with no
higher order terms. For this class of models $\fnl=0$ to all orders
in perturbation theory, within the same patch that this ansatz for $\zeta$ is valid in. At tree-level $\taunl$ also vanishes,
$\taunl^{\rm tree}=0$, but the unique one-loop corrections generate a
non-vanishing result, $\taunl \approx (54/25)^2\gnl^2{\cal
P}_{\zeta}{\rm ln (k\ell)}+{\cal O}(\gnl^3{\cal
P}^2_{\zeta})\lesssim 10^4$. There is no two-loop correction and the three-loop correction is constrained to be much less than unity due to the observational bound on $\gnl$. The relation $\taunl = (6\fnl/5)^2$ is
therefore clearly violated by the loop corrections. To the best of
our knowledge, this has not been demonstrated previously. However
the Suyama-Yamaguchi (SY) inequality, $\taunl \geqslant
(6\fnl/5)^2$, which was originally proved at tree level in
\cite{Suyama:2007bg}, and the possible effects of loop corrections
were discussed in
\cite{Suyama:2010uj,Sugiyama:2011jt,Bramante:2011zr}, has recently
been shown to remain true against loop corrections at all orders in
a completely model independent sense \cite{Smith:2011if} (although
there are subtelties about the definition of the local correlators
beyond tree level \cite{Bramante:2011zr}).

The previous example is by no means a special case: for generic
single-source models we also find that the relation $\taunl =
(6\fnl/5)^2$ can easily be broken by loops. Assuming there are no
large loop corrections with higher than three-point vertices, and
concentrating on the limit $|\gnl|\gg \fnl^2$, the one-loop
corrected $\fnl$ is given by $\fnl\,\approx\, \fnl^{\rm tree}(1+
13\,\gnl\,{\cal P}_{\zeta}\,\,{\rm ln}(k\ell))$. Using that
$|\gnl|\lesssim 10^{6}$, we conclude that the loop corrections to
$\fnl$ are small, at most at one per cent level, since $13\,\gnl\,
{\cal P}_{\zeta}\,\,{\rm ln}(k\ell)\,\lesssim \,10^{-2}$. On the
other hand, the one-loop corrections to $\taunl$ of Eq.
(\ref{taunl_loop_single}) can be large, significantly altering the
tree level relation $\taunl^{\rm tree} = (6\fnl^{\rm tree}/5)^2$.
Combining the results so far we find the general one-loop corrected
single-source relation between $\taunl$ and $\fnl$ given by
  \beq
  \label{taunl_fnl_single}
  \taunl\approx
  \left(\frac{6}{5}\right)^2\fnl^2+\left(\frac{54}{25}\right)^2\gnl^2\,{\cal P}_{\zeta}\,{\rm ln}(k\ell)
  %\,\left(1+{\cal O}(\gnl {\cal P}_{\zeta})\right)
  \ .
  \eeq
The second term on the right hand side arises from the loop
corrections and can be as large as $10^4$, irrespectively of the
value of $\fnl$. It may therefore give the dominant contribution to
$\taunl$.  In conclusion, the  condition  that leads to sizeable
inhomogeneities for $\fnl$, that is having a large $\gnl$, also
leads to breaking the single-source consistency relation between
$\fnl$ and $\taunl$.
 On the other hand, even though the loop corrections can cause
large deviations from the tree-level single-source result
$\taunl^{\rm tree} = (6\fnl^{\rm tree}/5)^2$, the relation between
$\taunl$ and $\fnl$ still serves as a useful discriminator between
single and multiple source models. Indeed as we will show in section
IV, the result (\ref{taunl_loop_single}) is converted  into a lower
bound for $\taunl^{1-{\rm loop}}$ in the multiple source case, which
also implies that (\ref{taunl_fnl_single}) becomes  a
lower bound for $\taunl$. % A new inequality can then be written,
 It would be interesting to generalize this analysis
allowing for large loop corrections with higher than
three-point vertices and see if the same conclusion holds even in
that case.

%     This effect  is expected to depend only  logarithmically on the size of the patches  \cite{Babich:2004yc,Riotto:2010nh}.  We will not further develop this issue in this  work, but to be conservative
%      we will increase of one order of
%     magnitude the values of our expected  constraints on higher order $n$-point  functions in the full sky {\bf CB we should be able to improve this argument soon}. {\bf GT indeed: we can write the equation for the S/N ratio, and briefly comment  that
%     on the other cases (trispectrum) something similar is expected}. We will see that, even in this case, we can find interesting constraints on these quantities.

% {\bf (SN: This doesn't seem to give any non-trivial bound on $f_{5}^{(3)}$.
%>From the loops we get a contribution ${\cal P}_{\zeta}f_{5}^{(3)}$ to $\fnl$ and requiring this to be smaller than
%$10^{2}$, we get basically the same bound on $f_{5}^{(3)}$.) CB, the loop you find is a "`dressed vertex", right? ie a 4th derivative and 2 first derivatives. These can perhaps be renormalised, lots of people do so, often citing my diagrammatic paper with David, Misao and Kazuya but ive never really understand if this physically works myself!}

\subsection{Correlators of four-point functions}\label{sec:4}

Considering correlators involving four-point functions measured in small patches, we can   constrain higher-order $n$-point
functions in the full sky, with $n\ge 5$. In
 the single-source case, the only new parameter which we can independently  constrain (even if going  beyond 2-point correlators) is the
 quantity
%  {\bf(SN: I think the numerical factor should be $\frac{1}{4!}\left(\frac{5}{3}\right)^3$: $\fnl=\frac{1}{2!}\frac{5}{3}\times$,
 %$\gnl=\frac{1}{3!}\left(\frac{5}{3}\right)^2\times$ etc.) }
 \be \hnl\,\equiv\,
 \frac{1}{4!}\left(\frac{5}{3}\right)^3
 \frac{N''''}{N'^4}\,. \ee
 This is associated with the fourth order term in the standard expansion of the curvature perturbation
 \be \zeta=\zeta_G+\frac35\fnl\,\zeta_G^2+\left(\frac35\right)^2\gnl\,\zeta_G^3 +\left(\frac35\right)^3\hnl\,\zeta_G^4+\cdots\,. \ee
 In multiple-source models, more possibilities arise as we are going to  discuss in
 section \ref{sec:multi};
  in this subsection, we focus
 on correlators that specifically depend on $\hnl$. These involve measurements of $\gnl$ in small patches, and read~\footnote{Details of these calculations are given by taking the single-source limit of the general, multiple-source calculation which we will present
 in section \ref{sec:multi}.}
%
  %  Considering correlators
 %
% This is clear by considering (\ref{zeta-deriv})--(\ref{g-deriv}), and in fact it is only correlators involving $\gnl$, see (\ref{g-deriv}), which will provide a sensitivity to $\hnl$.
%
\bea  \langle\Delta_\ell \zeta ({\bf x}_1)  \Delta_\ell \gnl ({\bf x}_2) \rangle&\simeq &-\frac{1}{10}\, \hnl {\cal P}_\zeta\,.
\label{zetgnl}
\\
\langle \Delta_\ell \ln{\cal P} ({\bf x}_1) \, \Delta_\ell \gnl ({\bf x}_2)\rangle
&\simeq & \frac{1}{25}\,\fnl\,\hnl\,{\cal P}_\zeta  \\
\langle \Delta_\ell \fnl ({\bf x}_1) \, \Delta_\ell \gnl ({\bf x}_2)\rangle
&\simeq & \frac15\,\gnl\hnl\,{\cal P}_\zeta   \\
\langle \Delta_\ell \gnl ({\bf x}_1) \, \Delta_\ell \gnl ({\bf
x}_2)\rangle &\simeq&\frac{1}{5}\, \hnl^2\,{\cal P}_\zeta
\label{autog4} \eea
where in the right hand side of each correlator we only include the
terms proportional to $\hnl$ and that are weighted by the largest
coefficient, under the assumption that the actual values of $\fnl$,
and $\gnl$ saturate their observational bounds. For simplicity, we
also drop the factor $F$ defined in Eq. (\ref{F}) which is of order
unity and does not modify our discussion.

Approximate $1\sigma$ constraints with WMAP for $\zeta,\,
\log {\cal P}_\zeta,\,
\fnl,\,\gnl$ are  respectively $10^{-5},\,10^{-1}\,,10 \,, 10^6$.
 Considering around one hundred
  patches of the sky, the measurement errors in each patch will be about ten times larger than these.
 On the other hand, when making correlations over the full sky, we should be able to recover the accuracy
 we loose when focussing on a small patch.  As discussed
at the end of subsection \ref{sub-errors},  in order to  take into account
 effects associated with the fact that only a limited class of momentum space
 configurations fits into the small patches, we increase
  the expected error bars on the  correlations above  by one order of magnitude. The resulting error bars are
consequently of order
  $10^2,\,10^6,\, 10^{8}$ and  $10^{13}$, respectively for eqs. (\ref{zetgnl})--(\ref{autog4}). Considering the constraints these terms could give on $\hnl$, we see the  correlation between $\zeta$ and $\gnl$,
 eq. (\ref{zetgnl}),  would lead to a competitive constraint, of
  order
  \be \label{chnl}
  | \hnl |\lesssim {\rm few}\times 10^{11}\,.\ee
  Moreover, this correlation is sensitive to the sign of $ \hnl$. A comparable constraint can be
  obtained from the  autocorrelation of $\gnl$,  eq. (\ref{autog4}),  although this quantity cannot determine the sign
  of $\hnl$~\footnote{  Planck data will be characterized by error bars  reduced by one order of magnitude
 with respect with the ones discussed above. The very  same procedure could then be applied,
 obtaining more stringent limits on $\hnl$.
 }.

Note that the assumption $| \hnl |\ll 10^{5}|\gnl|$, which justifies
neglecting the second order terms in eq. (\ref{fnl_expand}), breaks
down when $|\gnl|\sim 10^6$ and the bound $| \hnl |\lesssim {\rm
few} \times 10^{11}$ is saturated. To work out the precise numerical
factors in this limit one should therefore include the second-order
terms and repeat the analysis leading to eqs. (\ref{zetgnl}) -
(\ref{autog4}). This however should not affect the order of
magnitude of the bound $| \hnl |\lesssim 10^{11}$.

% for quantities

 %discussed above by on
%\bigskip

How useful is this constraint on $\hnl$? As we did in the previous
sections with respect to $\gnl$, we should compare
them with constraints obtained from
 loop contributions to inflationary observables that
  are sensitive to $\hnl$.
%  we already discussed in some length
  %with respect to inhomogeneities of $\fnl$. With respect to the possibility of constraining
  %$\hnl$,
 There is a
``dressed-vertex" loop diagram of the bispectrum \cite{Byrnes:2007tm} which gives
$\fnl^{1-{\rm loop}}\sim \hnl P_\zeta$, which gives effectively a constraint of the same order as
(\ref{chnl}), unless there is a cancellation between
loop terms and the tree level term contributing to  $\fnl$. One would need to
include numerical factors to see which constraint is the most competitive.

If one
renormalises the $\delta N$ coefficients to avoid the dressed
vertices, as suggested in \cite{Byrnes:2007tm}, one should shift the derivative of $N$ introducing
$\tilde{N}''=N''+N''''\langle\delta\phi^2\rangle/2$, which absorbs
this term. On the other hand, in this way, the fine-tuning needed  to compensate the
loop corrections does not
 go away, it just
becomes harder to spot.   If the loops associated with dressed
vertices do cancel, then the largest loop term involving $\hnl$
becomes $\gnl^{1-{\rm loop}}\sim \fnl\hnl P_{\zeta}$, which implies
$|\hnl| \lesssim 10^{14}$, which is a much weaker constraint than we forecast
cound be found by considering the autocorrelation of $\Delta_l\gnl$.

\section{Multiple source models}\label{sec:multi}

The extension of the previous
 analysis to the multiple-source case
  suggests that our method is able
  to probe many of the parameters controlling
   higher-order point functions.

Taking ensemble averages of correlators
among $n$-point and $k$-point parameters calculated in
small patches, we are probing parameters associated to $(n+k)$-point
functions evaluated in the entire sky. More precisely, for $n\geq k$
the correlators of local $n$- and $k$-point functions probe those
parameters of the full-sky $(n+k)$-point function that contain at
most $n$:th order derivatives of $N$, the number of e-foldings. This
is the generalization of the observation made in \cite{Smith:2011if}
for the power spectrum. In  the single-source limit, some of the
$(n+k)$-point function parameters
  coincide with the squares of parameters associated
with lower point functions (analogously to the tree-level
consistency relation $\tau_{\rm NL} = \left( 6/5 f_{\rm NL}\right)^2
$ holding   between four and three point functions in the single
source case). This is the reason for which we found that the
two-point correlators of $\Delta_{\ell}f_{\rm NL}$'s calculated in
different small patches are proportional to $g_{\rm NL}^2$. On the
other hand, in the multiple source case, our method in principle
allows us to probe individually the various parameters controlling the
higher-point functions.

In this section, we discuss this topic
 in more detail.
Recently, a diagrammatic approach,  based on the $\delta
N$-formalism, has been adopted in
\cite{Suyama:2011qi,Yokoyama:2008by} to determine the parameters
characterizing the five and six point functions. While the
three-point function is characterized by one parameter $f_{\rm NL}$
and the four-point function by the two parameters $\tau_{\rm NL}$
and $g_{\rm NL}$, the five and six point functions are characterized
by three and six parameters, respectively. In this section, we
borrow the notation of \cite{Suyama:2011qi} for defining parameters
associated to $5$ and $6$-point functions
 (see  \cite{Suyama:2011qi} for more details, but notice that we do {\it not} adopt their normalization
 condition $N_c N_c =1$).
% higher order generalizations of the parameters $f_{\rm NL}$,
%$\gnl$, $\tnl$ in the multiple field case, and set inequalities among them.
%To write the extension of our results to multiple field case, we use their definitions
 %of higher order quantities.
%  We introduce the following parameters
 %(see  \cite{Suyama:2011qi} for more details, but notice that we do {\it not} adopt their normalization
 %condition $N_c N_c =1$)

 \bigskip
 \noindent
{\it Two \,\, point \,\, function:}
\bea
\noindent
% {\text \rm Three \,\, point \,\, function} \nonumber\\
P_{\zeta} = N_a N_a  \, \left( \frac{H}{2 \pi}\right)^2
\eea

 \smallskip
 \noindent
{\it Three \,\, point \,\, function:}
\bea
\noindent
% {\text \rm Three \,\, point \,\, function} \nonumber\\
\fnl = \frac{5}{6}\,\frac{N_a N_b N_{ab}}{\left( N_c N_c \right)^2}
\label{threepf}
\eea

 \smallskip
 \noindent
{\it Four \,\, point \,\, function:}

\bea
%\nonumber \\
%{\text \rm Four \,\, point \,\, function} \nonumber\\
 %\\
 \taunl&=& \frac{
N_a N_{a c} N_{ c b }  N_b }{\left( N_e N_e \right)^3} \hskip1cm
\gnl \,=\,\frac{25}{54} \frac{ N_a  N_b N_c  N_{a b c} }{\left( N_e
N_e \right)^3}\
\label{fourpf}
\eea

 \smallskip
 \noindent
{\it Five \,\, point \,\, function:}

\bea
%\\
 %{\text \rm Five \,\, point \,\, function} \nonumber\\
 f_5^{(1)}&=&\frac{N_a N_{a b} N_{b c} N_{c d } N_d}{\left( N_e N_e \right)^4}
 \hskip1cm
 f_5^{(2)}\,=\,\frac{N_a N_{a b} N_{b c d} N_{c } N_d}{\left( N_e N_e \right)^4}
 \hskip1cm
 f_5^{(3)}\,=\,\frac{N_a N_{ b} N_{c } N_{d } N_{a b c d}}{\left( N_e N_e \right)^4}
 \label{fivepf}
 \eea

 \smallskip
 \noindent
{\it Six \,\, point \,\, function:}

 \bea
\tau_6^{(1)}&=&    \frac{ N_a N_{a b} N_{b c } N_{c d } N_{d e} N_e
}{\left( N_f N_f \right)^5} \hskip0.6cm \tau_6^{(2)}\,=\, \frac{ N_a
N_b N_{a b c} N_{ c d e } N_{d } N_e }{\left( N_f N_f \right)^5}
\hskip 0.6cm g_6^{(1)} \,=\, \frac{ N_a  N_b N_{a b c} N_{ c d  }
N_{d e } N_e }{\left( N_f N_f \right)^5}\nonumber
\\
\label{sixpf}
g_6^{(2)} &=& \frac{
N_a  N_{a b} N_{b c d} N_{ c   } N_{d e } N_e
}{\left( N_f N_f \right)^5}
\hskip0.6cm
g_6^{(3)} \,= \, \frac{
N_a  N_{a b} N_{b c d e} N_{ c   } N_{d  } N_e
}{\left( N_f N_f \right)^5}
\hskip0.6cm
g_6^{(4)} \,=\, \frac{
N_a  N_b N_{  c} N_{  d  } N_{ e } N_{a b c d e}
}{\left( N_f N_f \right)^5}
\eea
Notice that the parameters associated with 3 and 4 point functions carry the conventional numerical
coefficients.
%In terms of usually defined quantities, one has for the three and four point
%function parameters
%\bea
%f_3 &=& \frac65 \fnl\hskip1cm
%\tau_4 \,=\, \tnl \hskip1cm
%g_4 \,=\, \frac{54}{25} \gnl
%\eea
The aim of this section is to show that ensemble averages of correlators
 between quantities involving two and three-point functions
  depend on parameters associated with five and six point functions.
  This suggests a method for probing the quantities appearing in
  eqs. (\ref{fivepf}) and
(\ref{sixpf}).

Generalizing the notation of the previous section, we can write in the multiple-field  case
 (summations over repeated indices are understood)
\be
 \fnl (\sigma^a_\ell, {\bf x})\,=\,\fnl
(\sigma^a_{H^{-1}})+f_{{\rm NL},\,b}
(\sigma^a_{H^{-1}})\,\Delta_\ell \sigma^b_\x
\ee
where the
lower-case upper indices $a,b,...$  denote the different scalar
fields involved. Analogous expansions hold for $\zeta$, ${\cal
P}_\zeta$, $\tau_{\rm NL}$ and $g_{\rm NL}$.
We assume that the metric in field space is flat, and  that
\be
 \langle \Delta_\ell \sigma^a_{\x_1} \Delta_\ell \sigma^b_{\x_2}
\rangle\,=\,\delta^{a b}\,\left( \frac{H}{2 \pi}\right)^2
\,\int_{k_{\text hor}}^{k_\ell}\,d k'\,\frac{\sin{(k'\,\Delta
x)}}{k'^2 \,\Delta x} \,.
 \ee
(We will neglect the
slow-roll suppressed $\sigma$-dependence of $H$ throughout this
paper).
Then we have
 \bea\label{zeta-deriv}
 \zeta_{,b} \,\Delta \sigma^b_{\x}&=& -\,N_b   \,\Delta
\sigma^b_{\x}\,,
\\
\left( \ln {\cal P_\zeta}\right)_{,\,b} \,\Delta \sigma^b_{\x} &=& 2\frac{ N_{a b} N_a}{N_c N_c}
\,\Delta \sigma^b_{\x}\,,
\\
f_{{\rm NL},\,d} \,\Delta \sigma^d_{\x}%&=&%\frac56
 %f_{3, \,a}\,\Delta \sigma^a(\bf x)\\
&=&\frac56
\left( \frac{2\,N_{a d} N_b N_{a b} }{ \left( N_{c} N_c\right)^2 }
+
 \frac{N_{a } N_b N_{a b d} }{ \left( N_{c} N_c\right)^2 }
 -
 \frac{24\, \fnl}{5}\,
 \frac{ N_e N_{e d} }{  N_{c} N_c }
\right)\,\Delta \sigma^d_{\x}\\
\tau_{{\rm NL},\,d} \,\Delta \sigma^d_{\x}&=&
2\,\left(\frac{N_{a d} N_{a c} N_{c b} N_b}{\left( N_e N_e \right)^3}
+\frac{N_{a } N_{a c d} N_{c b} N_b}{\left( N_e N_e \right)^3}
-3 \tau_{\rm NL}\,\frac{N_{c d} N_c}{N_e N_e}
\right)\,\Delta \sigma^d_{\x}\,,
\\
g_{{\rm NL},\,d} \,\Delta \sigma^d_{\x}&=&\frac{25}{54}
 \left(3 \frac{N_{a d} N_{b} N_{c } N_{a b c}}{\left( N_e N_e \right)^3}
+\frac{N_{a } N_{b} N_{c } N_{a b c d}}{\left( N_e N_e \right)^3}
-\frac{324\, \gnl}{25}\,\frac{N_{c d} N_c}{N_e N_e}
\right)\,\Delta \sigma^d_{\x}\,.\label{g-deriv}
\eea
Denoting
\be
\Delta_\ell \fnl ({\bf x})\,=\,\fnl (\sigma^a_\ell, {\bf x})-\fnl (\sigma^a_{H^{-1}})
\ee
and the same for $\zeta$, $\ln {\cal P}_\zeta$,  $\taunl$ and $\gnl$,
we obtain the following results (the function $F\equiv F(\Delta x, k_{\text{hor}}, k_\ell)$ is given in
Eq. (\ref{F})):
\bea \label{zetafnl}
 %\langle\Delta_\ell \zeta \Delta_\ell\ln P_\zeta\rangle&=&2f_3 P_\zeta F,
 %\\
  \langle\Delta_\ell  \zeta  ({\bf x}_1) \Delta_\ell \fnl ({\bf x}_2) \rangle&=& -\left(\frac53 \tau_{\rm NL}+\frac{9}{5} g_{\rm NL}-\frac{24}{5}\,\fnl^2\right){\cal P}_\zeta F\,,   \label{1-rel} \\
\langle\Delta_\ell \zeta ({\bf x}_1)  \Delta_\ell \taunl ({\bf x}_2)
  \rangle&=& -2\left(f_5^{(1)}+f_5^{(2)}-\frac{18}{5} \,\tau_{\rm NL} \fnl\right) {\cal P}_\zeta F\,,    \label{2-rel} \\
\langle\Delta_\ell \zeta ({\bf x}_1)  \Delta_\ell \gnl ({\bf x}_2)
\rangle&=& -\left( \frac{75}{54}\, f_5^{(2)}+\frac{25}{54}\,
f_5^{(3)}-\frac{36}{5}\,g_{\rm NL}\,\fnl\right) {\cal P}_\zeta F\,,
\label{3-rel}
\\
\langle \Delta_\ell \ln{\cal P} ({\bf x}_1) \,\Delta_\ell \ln{\cal P} ({\bf x}_2)\rangle
&=& 4 \tau_{\rm NL} \, {\cal P}_\zeta\,F\,, \label{tpps}
\\ \label{tpfnl}
\langle \Delta_\ell \fnl ({\bf x}_1) \,\Delta_\ell \fnl ({\bf
x}_2)\rangle &=& \frac{25}{36} \left(4 \tau_6^{(1)}+
\tau_6^{(2)}+\frac{576 \fnl^{2}}{25}\,  \tau_{\rm NL} +4
g^{(1)}_6-\frac{96 \fnl }{5} \,f_5^{(1)} -\frac{48  \fnl }{5}
f_5^{(2)}
\right)\,{\cal P}_\zeta\,F\,,%(\Delta x, k_{\text hor}, k_\ell)
\\
\langle \Delta_\ell \ln{\cal P} ({\bf x}_1) \, \Delta_\ell \fnl ({\bf x}_2)\rangle
&=&\frac56\,
\left(4 f_{5}^{(1)}+2 f_5^{(2)}-\frac{48}{5} \fnl \taunl
\right)\,{\cal P}_\zeta\,F\,,  \label{4-rel}
\\
\langle \Delta_\ell \ln{\cal P} ({\bf x}_1) \, \Delta_\ell \tau_{\rm NL} ({\bf x}_2)\rangle
&=& 4\,
\left(\tau_6^{(1)}+g_6^{(2)}-3 \tau_{\rm NL}^2
\right)\,{\cal P}_\zeta\,F\,,  \label{5-rel}
\\
\langle \Delta_\ell \ln{\cal P} ({\bf x}_1) \, \Delta_\ell \gnl ({\bf
x}_2)\rangle &=& \frac{25}{27}\, \left( 3
g_6^{(1)}+g_6^{(3)}-\frac{324}{25} \gnl \taunl \right)\,{\cal
P}_\zeta\,F\,.  \label{6-rel} \eea
These ensemble
averages are able to probe  combinations of parameters controlling 5
and 6 point functions, as explained above. The previous combinations
allow one to individually test {\it each} of the parameters characterizing the
five-point function. Indeed, a combination of Eqs. (\ref{2-rel}),
(\ref{3-rel}) and (\ref{4-rel}) lead to the following expressions
for these parameters evaluated in the full sky \bea f_5^{(1)}\,{\cal
P}_\zeta\,F &=& \frac35\, \langle \Delta_\ell \ln{\cal P} ({\bf
x}_1) \, \Delta_\ell \fnl ({\bf x}_2)\rangle +\frac12
\langle\Delta_\ell \zeta ({\bf x}_1)  \Delta_\ell \taunl ({\bf x}_2)
  \rangle+\frac{6}{5} \fnl \,\taunl \,{\cal P}_\zeta\,F\,, \\
  f_5^{(2)}\,{\cal P}_\zeta\,F &=&-f_5^{(1)}\,{\cal P}_\zeta\,F
  -\frac12 \langle\Delta_\ell \zeta ({\bf x}_1)  \Delta_\ell \taunl ({\bf x}_2)
  \rangle+\frac{18}{5} \fnl \, \taunl \,{\cal P}_\zeta\,F\,,\\
  f_5^{(3)}\,{\cal P}_\zeta\,F &=&-\frac13 f_5^{(2)}\,{\cal P}_\zeta\,F
  -\frac{54}{25} \langle\Delta_\ell \zeta ({\bf x}_1)  \Delta_\ell \gnl ({\bf x}_2)
  \rangle+\frac{1944}{125} \fnl\, \gnl \,{\cal P}_\zeta\,F\,.
  \eea
In order to individually test the quantities  associated with 6-pt
functions one can consider correlations among $\taunl$ and $\gnl$
measured in small patches -- although the expressions for these
quantities depend also on  7 and 8-pt functions. However  the parameter
$g_{6}^{(4)}$, containing fifth order derivatives, can not
be probed by considering the inhomogeneities of trispectrum to
leading order in $\Delta_{\ell}\sigma$. In order to probe
$g_{6}^{(4)}$, one should consider the inhomogeneities of the 5-point
function or include the next-to-leading-order contributions in
$\Delta_{\ell}\sigma$. However these  in general are suppressed by a further factor of ${\cal
P}_{\zeta}$.

 It is straightforward to check that, in the single-source limit, one recovers the results of the previous sections.
This is due to consistency relations associating parameters in
eqs
  (\ref{fivepf}) and
(\ref{sixpf})  with parameters in   (\ref{threepf}) and
(\ref{fourpf}). In the multiple source case, the parameters  of eqs.
(\ref{fivepf}) and (\ref{sixpf})  are not directly constrained by
the observational bounds on $\fnl$, $\taunl$ and $\gnl$. The
inhomogeneities we are calculating could therefore be very large in
certain classes of models -- even at the level that might be
detectable by re-analysing the WMAP data.
% See discussion in the next section.
 %A comment with respect the comparison between our results and \cite{Smith:2011if}:
 % our eq. (\ref{tpps}) corresponds to their eq. 6. You can appreciate that they coincide -- and
  %this's reassuring -- a part from
 % the order one log-enhancement that is absent in their expression.
 %  Possibly this due to
  % their 'trick' to integrate over the narrow band.

\section{When is the inhomogeneity of non-Gaussian parameters large?}\label{sec:large}

In this section, we investigate in more detail when the
autocorrelations of $\fnl$ and $\gnl$ measured in distinct patches
can be large, and discuss the differences between single and
multiple-source scenarios. We focus our attention on a
representative, two-field case, in which the curvature perturbation
is  expanded as
%
%{\bf GT to do: put right all the numerical coefficients }
%
%\subsection{Inhomogeneities of $\fnl$}
%
%
%
%
%
%We have seen that in single source models, a measurable inhomogeneity of $\fnl$ must be due to $\gnl\sim10^6$, which should also be detectable with Planck. We will heform re study under which additional circumstances the inhomogeneity may become large in multi-source models.
%
%In particular we will consider a two-field model. Since we may make a rotation in field space to the direction which generates the final curvature perturbation (which needn't be at all related to the adiabatic direction during inflation), we may parametrise
%
%$\zeta$ by
%
\bea\label{deltaNansatz} \zeta&=& N_1 \delta\phi+\frac12 N_{11}\delta\phi^2 +N_{12}\delta\phi\delta\chi +\frac12 N_{22}\delta\chi^2 + \frac16N_{111}\delta\phi^3+\frac12N_{112}\delta\phi^2\delta\chi +\frac12 N_{122}\delta\phi\delta\chi^2 +\frac16N_{222}\delta\chi^3\,. %+ N_{1 1 1 1} \,\delta\phi^4 + {\text{remaining terms with four derivatives}}
\eea
%
%Additional terms containing four derivatives of $N$ can be included, but will not play an important role  in what follows.
An expression of this form  can be obtained by doing a rotation  in
field space to the direction, $\phi$, generating the first-order
contribution to the power spectrum of $\zeta$; in this way only one
 linear term in scalar perturbations appear in Eq. (\ref{deltaNansatz}).
%
%We will soon see that the final two terms, which are placed in brackets, are irrelevant. %Although there may be two more terms at cubic order neither of them contribute to (\ref{tpfnl}), so we neglect them.
One could instead choose a different field basis in which the second
order derivatives are diagonalised (i.e.~$N_{12}=0$), as was chosen
in \cite{Sugiyama:2011jt}, but this would be at the expense of
introducing a second linear term which leads to more cumbersome
expressions for the non-linearity parameters. We note that, in
general,  either choice of eliminating one term from the $\delta N$
expression (\ref{deltaNansatz}) will lead to a non-trivial
field-metric, because the required field space rotation will depend
on the scale we are focussing in. However since we can only observe
a limited range of scales (around 5 e-foldings) this should
typically only introduce small corrections, unless there happen to be large numerical factors which override the expected slow-roll suppression.
%at least in models where
%the curvature perturbation is generated during slow roll inflation
%and remains frozen after the end of inflation.
We will not develop
this interesting issue further,  but simply adopt the expression
(\ref{deltaNansatz}) as a specific Ansatz for the curvature
perturbations.

Within this set-up, the
  non-linearity parameters controlling quantities up to the 6-pt function are:
\bea && \fnl=\frac56 \frac{N_{11}}{N_1^2}, \qquad f_5^{(1)}=\frac{1}{N_1^6}\left(N_{11}^3+2N_{11}N_{12}^2+N_{12}^2N_{22}\right), \qquad f_5^{(2)}=\frac{1}{N_1^5}\left(N_{11}N_{111}+N_{12}N_{112}\right), \\ && \taunl=\frac{1}{N_1^4}\left(N_{11}^2+N_{12}^2\right), \qquad \tau_6^{(1)}=\frac{1}{N_1^8}\left(N_{11}^4+3N_{11}^2N_{12}^2+N_{12}^4 +2N_{11}N_{12}^2N_{22}+N_{12}^2N_{22}^2\right), \qquad \\ &&\tau_6^{(2)}=\frac{1}{N_1^6} \left(N_{111}^2+N_{112}^2\right), \quad % \\ &&
\gnl\,=\,
\frac{25}{54}\frac{N_{111}}{N_1^3}, \quad
 g_6^{(1)}=\frac{1}{N_1^7}\left(N_{111}(N_{11}^2+N_{12}^2) +N_{112}N_{12}(N_{11}+N_{22}) \right)\\
 &&  g_6^{(2)}\,=\,\frac{1}{N_1^3}\left(
 N_{11}^2 N_{111}+N_{12}^2 N_{122}+2 N_{12} N_{11}N_{112}\right)\,,
\qquad
 g_6^{(3)} \,=\,
 \frac{1}{N_1^6}\left(
 N_{1111}N_{11}+N_{1112}N_{12}\right)\,. \eea
%Notice that the final two terms in (\ref{deltaNansatz}) do not appear, and their values are hence irrelevant for our purposes.

\subsection{Inhomogeneous   autocorrelation of $\fnl$}\label{sub_afnl}

We have learned that, in a single-source set-up, the autocorrelation
of $\fnl$ can be large provided that we saturate the observational
limit for the full sky value of $\gnl$. Let us investigate what
happens instead in the multiple source case. In order to get an
inhomogeneity for the parameter $\fnl$ large enough to be
potentially observable with Planck, we need the combination in
brackets of (\ref{tpfnl}) to be of order ${\cal O}(10^{12})$.

Given the present observational constraints,
$|\fnl|\lesssim10^2,\,\tau_{\rm NL}\lesssim10^4,\, |g_{\rm
NL}|\lesssim10^6$
\cite{Komatsu:2010fb,Desjacques:2009jb,Vielva:2009jz,Smidt:2010sv,Fergusson:2010gn,Smidt:2010ra},
one may check that the only term which could be so large is the one
proportional to $\tau_6^{(2)}$, that provides
\bea\label{maybe-large} \langle \Delta_\ell \fnl ({\bf x}_1)
\,\Delta_\ell \fnl ({\bf x}_2)\rangle &\approx& \frac{25}{36}\,
\tau_6^{(2)}{\cal P}_\zeta\,F \,=\,\left( \left(\frac{9}{5}
\gnl\right)^2 + \left( \frac{5}{6}\frac{N_{112}}{N_1^3}\right)^2
\right){\cal P}_\zeta\,F \,. \eea
%In deriving the above equation, all three non-linearity parameters have been assumed to saturate their observational bounds, which is required to make the magnitude of the above terms as large as possible.
Notice that the first term in the right hand side with $\gnl$ is the
same leading contribution we found in the single-source case, see
Eq. (\ref{dfnl_2point_complete}). The second term represents a pure
multiple-source contribution and its magnitude is independent of
$\gnl$. While the form of $\tau_{6}^{(2)}$ in Eq.
(\ref{maybe-large}) is specific for the two-source model
(\ref{deltaNansatz}) that we are considering as an example, the
conclusion that the inhomogeneities of $\fnl$ are not uniquely
determined by the magnitude of $\gnl$ holds for generic
multiple-source scenarios, as we will show below.

The result (\ref{maybe-large}) shows that large inhomogeneities of
$\fnl$ are generated by cubic derivatives of $N$, which also
generate loop corrections to $\taunl$. In the limit $\fnl^2\ll
|\gnl|$, the dominant part of one-loop corrections to $\taunl$ is
given by
  \bea \tau_{\rm NL}^{1-{\rm loop}}\approx
  \left(\tau_{6}^{(2)}+\frac{N_{112}^2}{N_1^6}+\frac{N_{122}^2}{N_1^6}\right){\cal
  P}_{\zeta}{\rm ln}(k\ell)\ ,
  \eea
which implies that
  \bea
  \label{multi_fnl2}
  \langle (\Delta_\ell \fnl)^2 \rangle\, \approx \left(\frac{5}{6}\right)^2\tau_{\rm NL}^{1-{\rm
  loop}}-\left(\frac{5}{6}\right)^2\left(\frac{N_{112}^2}{N_1^6}+\frac{N_{122}^2}{N_1^6}\right){\cal P}_{\zeta}{\rm ln}(k\ell)
  \leqslant \left(\frac{5}{6}\right)^2\tau_{\rm NL}^{1-{\rm
  loop}}\ .
  \eea
This has to be compared with what we found in the single-source
case, Eq. (\ref{inlop}), in which the autocorrelation of $\fnl$ is
uniquely determined by $\tau_{\rm NL}^{1-{\rm loop}}$, or
equivalently by $\gnl^2\,{\cal P}_\zeta$ according to the relation
(\ref{taunl_loop_single}). In the multiple-source case,
$\taunl^{1-{\rm loop}}$ instead sets only an upper bound on the
inhomogeneities of $\fnl$ and the actual level of inhomogeneities
could be quite different from $\taunl^{1-{\rm loop}}$.

It is straightforward to show that this conclusion holds for generic
multiple-source models, thus opening interesting possibilities for
discriminating between single and multiple-source scenarios.
%This conclusion can be cast in the form of an
%inequality, as the following arguments show.
% In our multiple source set-up an inhomogeneity of $\fnl$ is always
%due to a cubic derivative of $N$ which must also give a large
%contribution to the trispectrum. In fact,  $\tau_{\rm NL}^{1-{\rm
%loop}}\gtrsim \gnl^2{\cal P}_{\zeta}$ in generic models and large
%inhomogeneities therefore necessarily require a large $\taunl$
%generated by loops.
% Let us then discuss in more detail the size of loop corrections to $\taunl$ in this context, and
%how they can affect the inhomogeneity of $\fnl$.
Defining a unit vector $u^{(k)}_a = \delta_{ak}$, the Cauchy-Schwarz
inequality for the inner product of the vectors
$u_{b}^{(k)}N_{a}N_{abc}$ and $N_{c}$ leads to
  \beq
  \label{cs_ineq}
  (u_{b}^{(k)}N_{a}N_{abc}N_{c})^2\leqslant (u_{b}^{(k)}N_{a}N_{abc}N_{cde}N_{d}
  u_{e}^{(k)})(N_f N_f)\ ,
  \eeq
where only the repeated lower case indices are summed over. The
one-loop corrections to $\taunl$ containing two three-point vertices
are given by
  \beq
  \taunl^{1-{\rm loop}} = \frac{N_{a}N_{abc}N_{bcd}N_{d}}{(N_{e}N_{e})^4}{\cal P}_{\zeta}
  = \sum_{k}\frac{u_{b}^{(k)}N_{a}N_{abc}N_{cde}N_{d}
  u_{e}^{(k)}}{(N_f N_f)^4}{\cal P}_{\zeta}\ ,
  \eeq
where we have neglected the logarithm associated with loop
integrals. This represents the dominant contribution to the loop
corrections,  provided that $|\gnl|\gg \fnl^2$, and that there are
no large loops associated with higher than three-point vertices.
Similarly we can write
  \beq
  \tau_{6}^{(2)} = \frac{N_a  N_b N_{a b c} N_{ c d e } N_{d } N_e }{( N_f
  N_f)^5} = \sum_{k}\frac{(u^{(k)}_{b} N_a N_{a b c} N_c)^2}{( N_f
  N_f)^5}\ .
  \eeq

Comparing the above expressions to the inequality (\ref{cs_ineq}),
and making use of the inequality $
\left({54}/{25}\right)^2  \gnl^2
\,\leqslant\, \tau_{6}^{(2)}$
derived in \cite{Suyama:2011qi}, we find
  \beq
  \label{gnl_ineq}
  \left(\frac{54}{25}\right)^2{\cal P}_{\zeta}\gnl^2\leqslant{\cal P}_{\zeta}\tau_{6}^{(2)} \leqslant \taunl^{1-{\rm loop}}\ .
  \eeq
%Using the observational constraints and the inequalities of
%\cite{Suyama:2011qi}, one can show that magnitude of the
%inhomogeneities of $\fnl$ in equation (\ref{tpfnl}) is essentially set by the
%parameter $\tau_{6}^{(2)}$, at least  in the limit $|\gnl|\gg \fnl^2$.
%{(\bf
%SN: Actually one should also show that $\tau_{6}^{(1)} \leqslant$
%$\taunl$-loop proportional to $N_{ab}N_{bc}N_{cd}N_{da}$, which is
%easy, and argue that these loops are subdominant, which seems clear
%but appears harder to prove rigorously.)}
In the  limit  $|\gnl|\gg \fnl^2$, using Eq.  (\ref{tpfnl}) we
therefore find the general result
%{\bf GT improved the numerical coefficients: the previous version of the formula looked a bit misleading}
  \bea
  \label{multi_fnl22}
  \langle (\Delta_\ell \fnl)^2 \rangle
    \lesssim \left(\frac56\right)^2 \taunl^{1-{\rm loop}}\ .
  \eea
The inequality is saturated in the single-source case (\ref{inlop})
where the magnitude of the inhomogeneities is set by $\taunl^{1-{\rm
loop}}$. In multiple source scenarios, $\taunl^{1-{\rm loop}}$ can
be greater than $\langle(\Delta_{\ell}\fnl)^2\rangle$. The
inhomogeneities of $\fnl$ could therefore provide an interesting new
tool for discriminating between single and multiple source models.
While the result (\ref{multi_fnl22}) only applies in the limit
$|\gnl|\gg \fnl^2$ and assuming there are no large loops with higher
than three-point vertices, it could nevertheless offer an intriguing
new window for probing inflationary physics.

Similar information can be obtained by considering the structure of
the trispectrum. The inequality (\ref{gnl_ineq}) implies
  \beq
  \label{trispectrum_test}
  \left(\frac{54}{25}\right)^2{\cal P}_{\zeta}\gnl^2\leqslant
  \taunl^{1-{\rm loop}}\ ,
  \eeq
which is again saturated for the single-source case. Under the
assumptions stated above, there are no significant loop corrections
to $\gnl$, and $\taunl$ is dominated by the loop correction as we
are considering the limit $\tau_{{\rm NL}}^{\rm tree}\sim \fnl^2\ll
|\gnl|$. We therefore see that both the inhomogeneities of $\fnl$
and the ratio of the two parameters $\gnl$ and $\taunl$ have the
opportunity to provide interesting new tools for distinguishing
between single and multiple source models.

\subsection{Inhomogeneities of $\gnl$}\label{sec:gnl}

We can apply the same procedure to  analyze the autocorrelator of   $\gnl$.
This quantity is given by
\bea \label{corgnl} \langle \Delta_\ell \gnl ({\bf x}_1) \, \Delta_\ell \gnl ({\bf
x}_2)\rangle &=&  \left( \frac{625}{324}\tau_8^{(1)} +\frac{625}{2916}\tau_8^{(3)}-36\gnl^2\taunl
+\frac{3750}{2916}g_8^{(2)}-\frac{50}{3}\gnl g_6^{(1)} -\frac{50}{9}\gnl g_6^{(3)} \right)\,{\cal
P}_\zeta\,F\,. \eea
Based on a measurement accuracy of $|\gnl|\sim10^6$ (which will be
improved by up to two orders of magnitude with Planck
\cite{Smidt:2010ra}) we require the term in brackets to be ${\cal
O}(10^{22})$ in order to be able to probe the inhomogeneity with
WMAP data.  With Planck data values as small as ${\cal O}(10^{18})$
could be relevant.
%although the constraints on the non-linearity parameters will also become much tighter.

The new non-linearity parameters which enter at this order are:
\bea g_6^{(3)} N_1^6&=&N_{1111}N_{11}+N_{1112}N_{12}, \qquad \tau_8^{(3)} N_1^8=N_{1111}^2+N_{1112}^2,
\\  \tau_8^{(1)} N_1^{10}&=&N_{111}^2N_{11}(N_{11}+N_{12})+ 2N_{111}N_{112}N_{12}(N_{11}+N_{22}) +N_{112}^2(N_{22}^2+N_{22}^2), \\ g_8^{(2)} N_1^9&=& N_{1111}(N_{11}N_{111}+N_{12}N_{112})+N_{1112}(N_{12}N_{111}+N_{22}N_{112})\,.
\eea
Notice that only two new quantities are important for studying the
new correlator of Eq. (\ref{corgnl}). These
 are $N_{1111}$ and $N_{1112}$, which are two out of the five fourth derivatives of $N$. Unless one of these two quantities is large, the correlator (\ref{corgnl}) does not give competitive information compared to the correlators  we considered in the previous sections. If one of the fourth derivatives is assumed to be very large, then the dominant  term is simply $\tau_8^{(3)}$. This implies that  we can constrain $N_{1111}/N_1^4$ and $N_{1112}/N_1^4$  to the level of $10^{15}$ with WMAP and $10^{13}$
with Planck.

On the other hand, $|\fnl^{1-{\rm loop}}|\simeq |N_{1111}|/N_1^4
P_\zeta \lesssim 10^2$ provides the constraint
$|N_{1111}|/N_1^4\lesssim10^{11}$, so barring an accurate
cancellation between this term and the tree level $\fnl^{\rm tree}$
we do not find interesting constraints from this contribution.
However for $N_{1112}$ there is no 1--loop constraint from $\fnl$,
and instead the tightest constraint comes from $\tau_{\rm
NL}^{1-{\rm loop}}=N_{1112}N_{12}P_\zeta /N_1^6\lesssim10^4$. This
does not constitute a  dominant constraint compared to the expected
sensitivity with the Planck satellite unless $|N_{12}|/N_1^2$ is
larger than unity.
%; for comparison purposes, one gets that $|\fnl|\lesssim10^2$ implies $|N_{11}|/N_1^2\lesssim10^2$.

In conclusion, we see some similarities between studying the autocorrelation of $\Delta_\ell\fnl$ and here the autocorrelation of $\Delta_\ell\gnl$. In both cases one can constrain two higher order derivatives, the one which is present in the single-source case, i.e.~$N_{111}$ and $N_{1111}$ which are proportional to $\gnl$ and $\hnl$ respectively, but this constraint is not very competitive compared to the natural  constraint we get from consider the loop contribution to lower order correlators unless there is a chance cancellation between the loop and the tree level terms. For the `cross-derivatives', $N_{112}$ and $N_{1112}$, the loop constraints are weaker (especially in the case of $N_{112}$) and we can achieve a tighter probe by considering the inhomogeneity of non-Gaussian correlators.

\section{Conclusions}\label{sec:conclusions}

In this paper, we investigated under which conditions  inflationary parameters can be inhomogeneous to an  observable level,
 focussing in particular on observables controlling local non-Gaussianity.
 We have demonstrated  that if we subdivide the entire  sky into a large number of small patches, the value of cosmological observables
 associated with the properties of the curvature fluctuation
can differ from patch to patch.
%  Indeed,  each small part of the sky experiences a different cosmological background
%over which the curvature fluctuation is defined,
%due to the cumulative effect  of  random modes with wavelengths larger  than the size of each patch.
 In particular, we have
shown that
correlators between $n$-point functions of curvature fluctuations as measured in one patch, and
$k$-point functions as measured in another patch, depend on $(n+k)-$point functions in the entire sky. This implies
that the expected degree of inhomogeneity in observable quantities  can be quantified
in a rather model independent manner. In interesting cases it
 results large enough to be measurable,
  since at present we have only weak constraints on parameters controlling higher-order
point functions in the full sky. Consequently, inhomogeneities of non-Gaussian parameters
can also be seen  as feasible method   for probing or constraining higher-point functions.

\smallskip

We have analyzed in detail the degree of inhomogeneity of local non-Gaussian observables, first in the single-source
 case, in which only one scalar field contributes to the generation of primordial curvature perturbation (as in
 curvaton models),
 then in   multiple-source  set-ups. In the case of single-source models, we have shown that autocorrelators of
  $\fnl$ evaluated in different patches depend  on the value of $\gnl$ in the full sky. If $\gnl$ turns out to be large
  enough to saturate its present day bound, we should expect variations of $\fnl$ of order one hundred from patch to patch,
  large enough to be observable.  Autocorrelators of $\gnl$, on the other hand, can  be used to set constraints
  on $\hnl$, the parameter characterizing 5-point functions: present day  data are accurate enough to be
  able to set an upper bound
  $|\hnl| \,\lesssim \,{\rm few} \times 10^{11}$. In the case of multiple source models, we have shown that correlators
  between parameters controlling $n$-point functions in different patches, with $n\,\le\, 4$, can be useful
  for probing individually the several
  new parameters that characterize  five and six point functions.

\smallskip

We have pointed out interesting connections between the degree of inhomogeneities  and
loop corrections to  non-Gaussian observables. Typically, models that lead to sizable inhomogeneities
are also characterized by large loop corrections to  inflationary observables. We have used  this fact to determine
 consistency relations between quantities denoting respectively  inhomogeneities  and loop contributions
 to non-Gaussian  observables. These consistency relations take the form of inequalities in  multiple
 source models, that are saturated in the single-source limit: consequently, they offer new observational perspectives
   for
  distinguishing between single and multiple source set-ups.

 \smallskip

 The conclusion of  our  theoretical analysis is that a
 sizable degree of inhomogeneity in  non-Gaussian observables
 is allowed  by
 present day bounds  on $\gnl$ and $\taunl$.
 We have then described  observational prospects  for probing inhomogeneities of non-Gaussian observables, discussing
 the accuracy we should expect for  determining correlators  of $n$-point functions measured in
 different patches of the skies. We discussed  geometrical
   effects that reduce the accuracy, and conservatively took them into account in our estimates. We have also pointed out
   that alternative techniques based on needlets analysis of CMB data are particularly well suited for testing non-Gaussianity
   in selected regions of the sky. It would be very interesting to apply those techniques to study inhomogeneities.

\acknowledgments We thank Emma Beynon, Paolo Creminelli, Rob
Crittenden, Dominic Galliano, Alan Heavens, Antony Lewis, David Seery and Yuko Urakawa for discussions.  GT is supported by an STFC
Advanced Fellowship ST/H005498/1. DW is supported by STFC grant
 ST/H002774/1.
 CB and SN thank the ICG, University of
Portsmouth for hospitality on a visit during which this project was
initiated.

\end{document}